# Significance of the dispersion force for ferroelectric switching in ZnO and related materials

*Lingyao Zhang[1,2], Musen Li[1,3,4], Nisha Metha[3], Carla Verdi[5], Wei Ren,[1,4*] and Jeffrey R. Reimers[1,3*]*

1 Physics Department, State Key Laboratory of Advanced Refractories, Materials Genome Institute, International Centre for Quantum and Molecular Structures, Shanghai University, Shanghai 200444, China

2 Department of Materials Science, University of Milan-Bicocca, Via Roberto Cozzi 55, 20125 Milan, Italy

3 University of Technology Sydney, School of Mathematical and Physical Sciences, Ultimo, New South Wales 2007, Australia.

4 Institute for Quantum Science and Technology, Shanghai Engineering Research Center for Integrated Circuits and Advanced Display Materials, Shanghai University, Shanghai 200444, China.

5 School of Mathematics and Physics, The University of Queensland, Brisbane, Queensland, Australia

ABSTRACT: Wurtzite-ZnO is a wide-bandgap polar material with a ferroelectric-switching barrier that is too high to utilize, but the barrier can be reduced and switching observed in substituted materials such as $Zn_{0.5}Mg_{0.5}O$. Here, we seek to understand atomic-scale features that control concerted polarization switching in these and related systems, focusing on the planar hexagonal structures h-ZnO and $Zn_{0.5}Mg_{0.5}O$ that may act as metastable intermediate phases along the switching pathway. Consensus is obtained by considering a range of pure and dispersion-corrected density-functional theory (DFT) computational approaches, as well as *ab initio* Hartree-Fock (HF), Møller-Plesset perturbation-theory (MP2), and random-phase approximation (RPA) calculations. The perceived stability of h-ZnO is found to be strongly influenced by the dispersion correction, with the consensus being that dispersion interactions are insufficient to stabilize h-ZnO as a metastable phase in infinite crystals. In contrast, h-$Zn_{0.5}Mg_{0.5}O$ is consistently predicted to be at least metastable, with some dispersion-corrected DFT approaches predicting it to be more stable than its wurtzite form; all DFT methods overestimate its stability compared to MP2 and RPA. Dispersion forces are found to be most significant for hypothetical planar hexagonal structures constrained to the lattice vectors of the wurtzite phases. In general, our results demonstrate that an accurate treatment of dispersion forces is essential when describing polarization switching and ferroelectric behavior in wurtzite-structured materials.

1. **INTRODUCTION**

The properties of molecules, materials, and biological systems can be broadly understood as being controlled by three primary effects: (i) ionic bonding, a classical phenomenon in which charge localizes on certain atoms to form ions, maximizing their Coulomb interaction energy;[1,2] (ii) covalent bonding, a quantum effect in which electrons are shared between atoms to minimize their kinetic energy[3-6] and recognizes that fermions such as electrons are indistinguishable; and (iii) the van der Waals dispersion force, a quantum-fluctuation effect that gives rise to long-range interactions owing to the inability to confine quantum particles such as electrons.[7-9] Atoms are often categorized as being either "hard" or "soft".[10,11] Hard atoms undergo strong ionic and/or covalent interactions, generating ground states that are well separated in energy from the excited states, whereas soft atoms often interact more weakly, are highly polarizable, and generate low-energy excited-state manifolds. Dispersion interactions are often neglected when considering materials composed of hard atoms as it is the polarizability that drives quantum fluctuations,[8,12] with an iconic example of this practice being zinc oxide (ZnO). Herein, we demonstrate that even for ZnO and related materials, the dispersion force is significant for the understanding of an important materials property: ferroelectric switching.

Polarization switching in wurtzite-structured materials such as ZnO (structure **A** shown in Fig. 1)[13] is of significant interest owing to the high bandgap and stability achievable within this materials class.[14-17] While wurtzite ZnO itself has a large barrier to switching, which to date has prevented the experimental realization of ferroelectricity, related materials with non-stoichiometric compositions close to $Zn_{0.5}Mg_{0.5}O$[17-23] have demonstrated switchable polarization and are therefore identified as promising candidates for new-generation device materials.[14-17,23] In such materials, the simplest way to consider ferroelectric switching is via a concerted mechanism that passes through an unpolarized planar hexagonal intermediary structure **C** (Fig. 1), known as h-ZnO and other related names.[14,24-30] Understanding the properties of this and related structures is therefore critical to future developments in this field.[17,22,25-30]

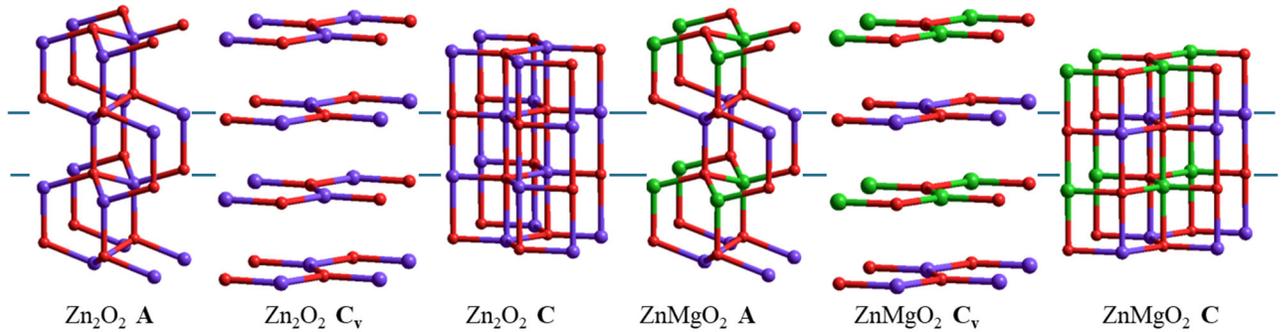

**Figure 1.** 2 × 2 × 2 replicas of the 4-atom unit cells of ZnO and $Zn_{0.5}Mg_{0.5}O$ in their wurtzite polymorphs **A** (tetrahedral coordination), their associated hexagonal planar (h-) polymorphs **C** (trigonal-bipyramidal coordination), and their hypothetical hexagonal planar structures constrained to the lattice vectors of **A**, denoted $C_v$ (trigonal planar coordination); red- O, green- Mg, purple- Zn.

In the wurtzite structure, the coordination about both the metal atoms and the oxygen atoms is tetrahedral, whereas in the planar hexagonal structure the coordination is trigonal bipyramidal (Fig. 1). The planar structures are nonpolar and have zero dipole moment, whereas in the wurtzite form, all metal atoms are displaced out of plane in one direction and the oxygen atoms in the opposite, creating a dipole polarization with two possible orientations.

For the parent material ZnO, layered thin films and nanostructures embodying h-ZnO have been stabilized on surfaces.[27,31] Their unique properties have led to proposals for their use in diverse fields including hydrogen storage[32] and thermovoltaics.[33] h-ZnO has also been reported in high pressure environments,[29,30,34-38] and the synthesis of high-purity h-ZnO nanocrystals that are stable under ambient conditions up to 200 °C has been reported.[24] However, the experimentally determined interatomic distances and lattice parameters of these crystals were well outside the ranges predicted by density functional theory (DFT) calculations,[29,30,34,39-41] as well as by *ab initio* Hartree Fock (HF) theory.[42] Moreover, computational studies have predicted h-ZnO to be a transition state for concerted ferroelectric switching, contrary to its experimental report as a crystalline metastable (intermediary) phase. In contrast, calculations of multiple types have supported the observation of supported h-ZnO layered films,[43-46] in agreement with experiment.

Hence, the existence of h-ZnO as a metastable phase under ambient conditions remained controversial until a sophisticated re-refinement[47] of the original experimental data[24] delivered structural parameters in good agreements with computational predictions. The main remaining issue is reconciliation of the computational results predicting that h-ZnO crystals are unstable with the experimental observation of metastable h-ZnO nanocrystals. A reason for this could be that calculations were performed for infinite crystals, whereas the observed nanocrystals may be sufficiently small for solvent effects or other environmental factors to influence crystal growth and stability. Another reason could be that the computational treatments applied to date were inadequate for the problem under consideration. Herein, we focus on this second possibility, seeking to obtain consistent qualitative features using a wide variety of computational approaches and hence reveal the importance of different effects to the switching process, as well as delivering the quantitative accuracy needed to aid the interpretation of the experimental data.

We start by noting that all previously applied computational approaches assume that the dispersion force could be neglected, owing to both Zn and O being hard atoms. This could be significant as the switching process involves large changes in the inter-layer $c$ lattice parameter (visually apparent in Fig. 1), and if h-ZnO or h-$Zn_{0.5}Mg_{0.5}$O could be prepared at the lattice dimensions of their wurtzite forms, then the coordination would be trigonal planar (graphite or h-BN like)[22] rather than trigonal bipyramidal. These hypothetical structures, denoted $C_v$ in Fig. 1, suggest that dispersion interactions could make significant contributions to the broad understanding of ferroelectric switching. Nevertheless, calculated barriers for ferroelectric switching in these materials are greatly enhanced when artificial strain is applied to preserve cell dimensions, a feature that would act to mitigate this effect.[17,22,48,49]

The simplest level of *ab initio* theory used to model the properties of molecules and materials is HF theory,[50] which embodies ionic and covalent bonding but ignores the dispersion force. In HF theory, each electron is assumed to interact only with the mean-field of the other electrons, ignoring electron correlations that allow these equivalently charged particles to avoid each other better. A more sophisticated level of *ab initio* theory, second-order Møller-Plesset perturbation theory (MP2),[51] incorporates electron correlation beyond the mean-field level, as well as the related van der Waals dispersion force at an equivalent level. Extended to higher levels than MP2, the Møller-Plessett procedure, in principle, converges to the exact solution, although in practice this convergence is not guaranteed.[52,53] As MP2 is a perturbative correction to HF, it is only appropriate for materials with large band gaps, where the HF reference provides a qualitatively correct description, as indeed is the case for ZnO and $Zn_{0.5}Mg_{0.5}O$. A practical criterion for the appropriateness of MP2 is that the HF results should provide a realistic qualitative description of the phenomenon of interest that is simply just improved by MP2.

Typically, MP2 is known to overestimate the effects of electron correlation to the covalent bonding when considering reaction barrier heights,[54] dative covalent bonding[55] as found in ZnO and $Zn_{0.5}Mg_{0.5}O$, as well as the effects of dispersion.[56,57] If all correlation effects were systematically overestimated, the exact result would lie between the HF and MP2 prediction, close to MP2, but in practice the differing magnitudes of errors associated with covalent bonding and dispersion make both the sign and magnitude of the total error difficult to predict.

We therefore also consider calculations based on the random-phase approximation (RPA).[58] This is an *ab initio* approach that includes electron correlation to infinite order in the ring diagrams and provides a more balanced treatment of chemical bonding and dispersion interactions.[59] For computational efficiency, it is implemented using many approximations which reduce its intrinsic scaling from $n^6$ to $n^4$. In this work, RPA is implemented empirically using PBE[60] to generate its starting orbitals, without explicit evaluation of singles contributions, following common practice. RPA is largely insensitive to band-gap size and is generally more accurate than MP2, although its shortcomings are more difficult to predict. RPA accounts for more electron correlation effects than MP2, encompassing additional contributions to both chemical bonding and dispersion interactions, e.g., contributions to the dispersion force beyond the London contribution.[9]

Alternatively, in DFT-based approaches, a mean-field design is used that differs from that applied in HF theory so as to capture key aspects of the electron-correlation contributions to covalent bonding.[61,62] Its results are generally superior to those obtained from HF calculations, but exceptions can occur if the DFT functional describes charge-transfer sufficiently poorly to misrepresent the ground state[63] or excited state[64,65] of interest, although this scenario is not relevant for ZnO and $Zn_{0.5}Mg_{0.5}O$.

DFT is commonly implemented using the simple local-density approximation (LDA), the general-gradient approximation (GGA) which further improves the description of electron correlation, meta-GGA functionals that additionally refine the description of the electronic kinetic energy,[66] or hybrid functionals[67] that instead enhance the description of the electron exchange contribution. Examples of GGA-type functionals that have been applied to ZnO include PBE,[60] while examples of hybrid functionals include B3LYP,[67] PBE0,[68] and HSE06.[69,70] Herein, we perform calculations using a modification of PBE optimized for solids, PBESol,[71] the hybrid functionals HSE06, and PBE0, as well as the meta-GGA functional r$^2$SCAN.[72] Other enhanced approaches not considered here include asymptotically corrected hybrid functionals that allow for better descriptions of charge-transfer processes,[65,73-75] functionals that combine hybrid and meta-GGA properties,[76] and double-hybrid functionals that include MP2 contributions within the density functional.[77-81] When HF provides a realistic description of some phenomena, MP2 often outperforms GGA and hybrid DFT approaches, even when dispersion corrections are applied.[56,78,79]

Indeed, a shortcoming of all modern density functionals is that they do not intrinsically include the dispersion force that acts at long range both within and between materials and molecules. Of particular relevance here, the dispersion force has been shown to control ferroelectric switching in some susceptible materials.[82,83] To deal with this, many empirical dispersion corrections have been developed to enhance DFT functionals. In this work, we employ the D3(BJ)[78] correction for HF, PBESol, PBE0, and HSE06, with the rVV10[84] correction applied[85] for r2SCAN; some test calculations are also performed using other approaches.[86-97] Most dispersion corrections, including D3(BJ), account only for the London dispersion term, as does MP2. Other approaches are available that go beyond this[92,98] and can account, e.g., for the switching off of the dispersion force owing to Faraday-cage screening,[99] as does RPA. The most advanced modern density functionals, such as double-hybrid functionals, embody MP2-type correlation effects and typically outperform MP2 for chemical reactivities and dispersion interactions,[77,79-81] but remain of limited availability in computational packages for materials science. The MP2 method itself can also be improved by adding empirical dispersion corrections to mitigate its tendency to overestimate dispersion contributions.[56,100]

When HF provides a reasonable qualitative description of the phenomenon of interest, then MP2 offers a balanced systematic way of improving its description of ionic binding (charge transfer), covalent bonding, and dispersion interactions. Similarly, when PBE provides a reasonable qualitative description, our implementation of RPA should improve upon each of these aspects also in a balanced way. MP2 and RPA calculations are therefore expected to deliver the quantitative accuracy required to understand the nature of h-ZnO.

Used on their own, density-functionals provide (i) approximate treatments of the electronic kinetic energy that is critical to covalent bonding, (ii) either no treatment or empirical treatments of the asymptotic-potential effects associated with charge transfer and other aspects of the DFT self-interaction error, (iii) varying descriptions of short-range electron correlation, and (iv) either no treatment or empirical corrections for dispersion interactions. We consider the impact that these limitations have on polarization switching in ZnO and $Zn_{0.5}Mg_{0.5}O$, leading to a qualitative, energetics-based understanding of the ferroelectric-switching process and the stabilization of their critical planar-hexagonal metastable phases that is consistent with quantitative MP2 and RPA predictions.

## 2. METHODS

### 2.1 General

All calculations were performed using VASP,[101,102] Mostly, projector-augmented wave (PAW) pseudopotentials[103] were used with 6 valence electrons for O, 2 for Mg, and 12 for Zn. Calculations were performed using HF[50] and MP2,[51] as well as the PBESol,[71] PBE0,[68] HSE06,[69,70] and r$^2$SCAN[72] density functionals, both with and without empirical dispersion corrections, to generate potential-energy surfaces (PES) for concerted polarization switching. Specifically, the D3 dispersion correction with Becke-Johnson damping,[78] D3(BJ) was added to HF and to all DFT functionals except r$^2$SCAN, to which rVV10[84,85] was added instead. Crystal structure optimizations were performed for all computational methods except MP2 and RPA, with some test calculations also performed using different DFT dispersion corrections without structure re-optimization.

In all DFT and HF calculations, the plane-wave energy cutoff was set to 520 eV. High-quality integration grids were employed ("PREC=ACCURATE"), the standard Fourier-transform grids were used in the evaluation of the Hartree-Fock exchange ("PRECFOCK=NORMAL"), and projection operators were evaluated in reciprocal space ("LREAL=FALSE"). Gaussian smearing was used for the occupied orbitals with a width of 0.02 eV, $10^{-7}$ eV was used as the convergence criterion for the electronic-structure optimization, and $10^{-3}$ eV/Å as the force convergence criterion during structural optimizations.

### 2.2 Structure representation, optimization, and reaction coordinate definition

All structures were represented using unit cells containing four atoms, with compositions of either $Zn_2O_2$ representing ZnO and $ZnMgO_2$ representing $Zn_{0.5}Mg_{0.5}O$ (see Fig. 1). All reported energies pertain to these four-atom cells.

Structural optimizations were performed using a procedure intended to achieve high numerical precision, with the goal of reproducing total energies to within 0.1 meV. The primary optimization protocol was repeated 10 times to ensure robust convergence. In each repetition, the lattice vectors and all internal structural degrees of freedom were optimized simultaneously to convergence, using an initially-determined fixed plane-wave basis. As these plane-wave basis sets vary with the initial lattice size, subsequent calculations do not necessarily reproduce previous ones, leading to significant numerical noise in the calculated energies. The repetition strategy employed here reliably eliminates this effect.

The PES were constructed as functions of an internal reaction coordinate $RC$. This coordinate is defined in terms of the fractional displacement along the $c$-axis between the Mg atom in the $ZnMgO_2$ cell, or one of the Zn atoms in a $Zn_2O_2$ cell, and its bonded out-of-plane O atom. It is defined as the ratio of this displacement in an arbitrary structure to that in the optimized wurtzite structure **A**. Values of $RC = 0$ thus correspond to planar hexagonal structures **C**, whereas values of ±1 correspond the two symmetry-equivalent wurtzite structures **A** and **A′**. For $Zn_2O_2$, the symmetry is such that three variables determine the crystal structure: the lattice constants $a$ and $c$, and $RC$. For $ZnMgO_2$, two additional variables are required: the relative displacement of the O atom bonded to Zn from the Zn layer, and the asymmetry in the vertical positions of the Zn and Mg atoms. Calculations of the PES proceed by first optimizing **C** (two free variables) and **A** (three or five free variables for $Zn_2O_2$ and $ZnMgO_2$, respectively), followed by constrained optimizations at fixed values of $RC$ (two or four remaining free variables). Transition-state structures **B** and **B′** were identified as local maxima in the PES. Note that the calculation of phonon frequencies at constant volume is inappropriate for characterizing the nature of the PES in the vicinity of **C**, as the change in volume is an intricate aspect of the reaction pathway.

### 2.3 Brillouin-zone integration

The Brillouin-zone integrations were performed using Monkhorst-Pack $k$-point grids.[104] To ensure continuity of the PES, the same $k$-grid was used for all calculations, independent of unit-cell volume. Test DFT calculations performed using a denser $9 \times 9 \times 6$ $k$-point grid yielded results comparable to those obtained using a $6 \times 6 \times 4$ grid. As RPA is expected to be the method most sensitive to $k$-point sampling, a systematic convergence study was performed (see Supporting Information (SI)), with grids ranging from $3 \times 3 \times 2$ up to $12 \times 12 \times 8$. These tests likewise indicated convergence at $6 \times 6 \times 4$. Hence, this grid was used for all HF and DFT calculations.

### 2.4 MP2 calculations

Only the traditional $n^5$-scaling MP2 algorithm[105] was found to be suitable in this study, as more efficient schemes available such as Laplace-transform MP2 (LTMP2)[106,107] introduce numerical noise exceeding the limits required to obtain meaningful results. The traditional MP2 calculations included excitations into all unoccupied orbitals and used a maximum angular-momentum quantum number of 4 to treat the one-center terms.

To make traditional MP2 calculations computationally feasible in terms of both memory and CPU requirements, a three-step strategy was used, adjusting the plane-wave energy cutoff and the $k$-point sampling to meet differing challenges related to broad understanding and quantitative accuracy. First, MP2 calculations were performed for the key structures **A**, **C**, and ones relevant to **B**, using geometries previously optimized with each of the HF and DFT approaches, with the VASP-default plane-wave energy cutoff of 400 eV and a coarse $3 \times 3 \times 2$ $k$-point grid. This was used to identify a particular set of structures (those optimized with PBE0) to be used in all subsequent MP2 (and subsequent RPA) calculations. Second, PES were generated, for this chosen structure set, using an increased energy cutoff of 520 eV and the same small $k$-point grid. These results were then corrected by evaluating the corresponding DFT energy changes upon increasing the $k$-point grid to 6 × 6 × 4.  Third, a small number of MP2 calculations were performed for a critical subset of these structures, using the 520 eV cutoff and a 4 × 4 × 3 $k$-point grid, again applying DFT-based corrections to 6 × 6 × 4.  Test calculations indicated that the 4 × 4 × 3 grid captures over 90% of the DFT energy changes obtained when expanding the grid from 3 × 3 × 2 to 6 × 6 × 4.  Computational requirements for each of these MP2 calculations included approximately 1.7 TB of memory and 11,000 CPU hours.

### 2.5 RPA calculations

The RPA calculations[108] reported in the main text were mostly performed using a 6 × 6 × 4 $k$-point grid and a plane-wave energy cutoff of 520 eV cutoff , employing the less-approximate, full RPA implementation available in VASP with a 12-point frequency integration grid.  Some calculations were also performed for selected structures using the 6 × 6 × 4 $k$-point grid, an energy cutoff of 700 eV, a 24-point frequency integration grid, and "GW" pseudopotentials in place of the standard "PBE" ones.  This change in the pseudopotential increases the number of electrons included in the valence band to 10 for Mg and 20 for Zn.

## 3. RESULTS

An Excel file in SI lists detailed results for all calculations, including up to 46 optimized energetic properties, structural properties, key input parameters, and additional characterization data including convergence information. All mathematical manipulations of these data leading to the tables and figures presented in this work are also included therein. In addition, plots are provided, as function of the RC, for each energy component, the five geometrical variables, and the band gap. Extracted from this dataset, the calculated PES for polarization switching in $Zn_2O_2$ and $ZnMgO_2$ are shown in Fig. 2. This figure depicts the concerted switching pathway between the two symmetry-equivalent ferroelectric wurtzite structures, **A′** and **A**, via planar hexagonal structures **C**, possibly via intermediate transition states **B′** and **B**; properties evaluated at these critical structures are listed in Table 1.

### 3.1 Choice of optimized structures for the MP2 and RPA calculations

The MP2 and RPA calculations are performed at geometries optimised using DFT. This choice of DFT method to use was made based on two criteria: the structures that produced the lowest MP2 energies for **A**, **C**, and ~ **B**, and the structures providing the best agreement between calculated and experimentally observed lattice vectors. Results are listed in the SI, with the methods that predicted structures with the lowest MP2 energies arising from HF, PBE0, and $r^2$SCAN/rVV10, with the best lattice vectors produced by HSE06, PBE0, $r^2$SCAN, and $r^2$SCAN/rVV10 (see below). Although several methods performed comparably well, PBE0 was selected as it provided the most reliable description of the critical structure **B** of ZnO.

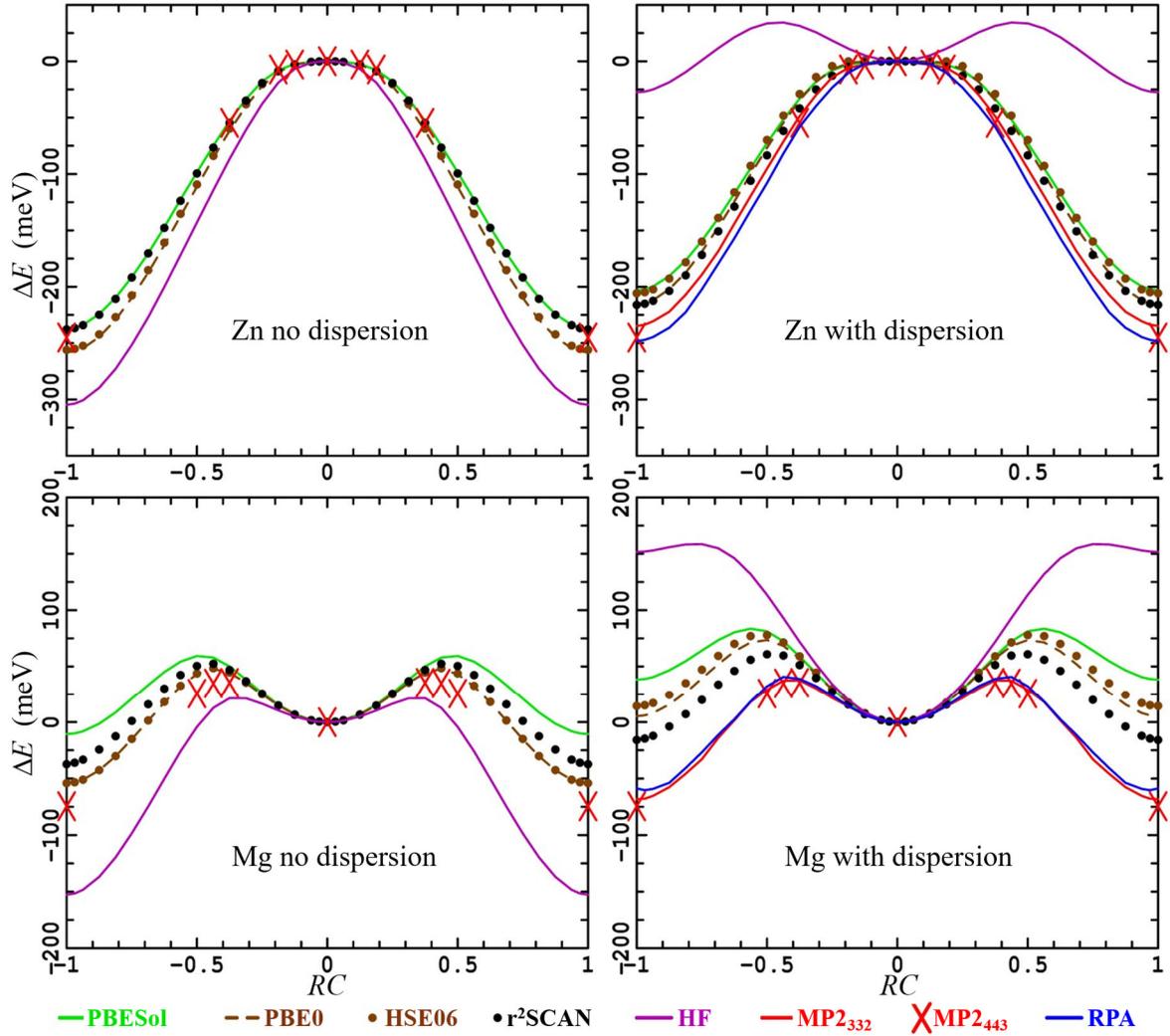

**Figure 2.** PES for concerted ferroelectric switching in unit cells of ZnMO$_2$, with M = Zn or Mg. The applied dispersion corrections are: that inherent to MP2 and RPA; D3(BJ) for HF, PBESol, PBE0, and HSE06; and rVV10 for r$^2$SCAN. Results pertain a 6 × 6 × 4 *k*-point grid, either by direct evaluation or else, for MP2$_{332}$ and MP2$_{443}$, the results are obtained by applying corrections to calculations performed using smaller grids as detailed in the text. Results are grouped according to methods with and without dispersion, with the MP2$_{443}$ results consistently displayed as a reference.

**Table 1.** Calculated energy differences between wurtzite structures **A**, planar hexagonal structures **C**, and possible[a] interconnecting transition states **B**, in meV per ZnMO$_2$ unit cell, obtained with (w D) and without (w/o D) contributions from dispersion.[b,e]

| M | Method | k-points | $E_C - E_A$ w/o D | $E_C - E_A$ w D | $E_B - E_A$ w/o D | $E_B - E_A$ w D | $E_B - E_C$ w/o D | $E_B - E_C$ w D | $E_{Cv} - E_A$ w/o D | $E_{Cv} - E_A$ w D |
|---|---|---|---|---|---|---|---|---|---|---|
| Zn | RPA[c] | 6 × 6 × 4[f] | - | 245 | - | - | - | -2 | - | 585 |
|  |  | 12 × 12 × 8 | - | 245 | - | - | - | -2 | - | 585 |
|  |  | 6 × 6 × 4 | - | 248 | - | - | - | -2 | - | 586 |
|  | MP2[c] | 4 × 4 × 3[d] | - | 245 | - | - | - | -3 | - | 593 |
|  |  | 3 × 3 × 2[d] | - | 235 | - | - | - | -1 | - | 589 |
|  | HF | 6 × 6 × 4 | 305 | 27 |  | 62 | -7 | 34 | 538 | 642 |
|  | PBESol | 6 × 6 × 4 | 236 | 203 |  |  | -2 | -1 | 551 | 588 |
|  | r²SCAN | 6 × 6 × 4 | 238 | 216 |  |  | -3 | -1 | 549 | 569 |
|  | PBE0 | 6 × 6 × 4 | 258 | 215 |  | 215 | -3 | 0.1 | 548 | 593 |
|  | HSE06 | 6 × 6 × 4 | 256 | 206 |  | 206 | -3 | 0.7 | 546 | 592 |
|  | CAM-B3LYP | 6 × 6 × 4 | 334 | 293 |  |  | *-18* | -7 | 514 | 567 |
| Mg | RPA[c] | 6 × 6 × 4[f] | - | 42 | - | 91 | - | 49 | - | 589 |
|  |  | 12 × 12 × 8 | - | 59 | - | 98 | - | 39 | - | 608 |
|  |  | 6 × 6 × 4 | - | 59 | - | 99 | - | 41 | - | 608 |
|  | MP2[c] | 4 × 4 × 3[d] | - | 74 | - | 109 | - | 35 | - | 596 |
|  |  | 3 × 3 × 2[d] | - | 68 | - | 106 | - | 38 | - | 594 |
|  | HF | 6 × 6 × 4 | 153 | -152 | 174 | -6 | 21 | 146 | 613 | 602 |
|  | PBESol | 6 × 6 × 4 | 11 | -38 | 70 | 45 | 59 | 84 | 529 | 538 |
|  | r²SCAN | 6 × 6 × 4 | 37 | 16 | 90 | 77 | 52 | 61 | 555 | 565 |
|  | PBE0 | 6 × 6 × 4 | 54 | -5 | 102 | 68 | 48 | 73 | 569 | 592 |
|  | HSE06 | 6 × 6 × 4 | 54 | -14 | 102 | 63 | 48 | 78 | 567 | 590 |
|  | CAM-B3LYP | 6 × 6 × 4 | 154 | 87 | 165 | 120 | 10 | 32 | 559 | 596 |

a: When no transition state is identified, the $E_B - E_A$ field is left blank and $E_B - E_C$ shows, in italics, the results for **B** at a reaction coordinate of ± 1/8; "-" indicates that results without dispersion are not possible for MP2 and RPA.

b: Either: the dispersion intrinsic to MP2 or RPA; D3(BJ) for HF, PBESol, PBE0, and HSE06; or rVV10 for r²SCAN.

c: Evaluated at PBE0 optimized structures.

d: A correction terms is added equal to that found using PBE0 when k-points are increased to 6 × 6 × 4.

e: Other calculations for M = Zn: $E_C - E_A$ by LDA 260 meV,[109] by PBESol 250 meV.[17] Other calculations for M = Mg: $E_B - E_A$ by PBESol ~80 meV,[17] $E_C - E_A$ by PBESol ~15 meV.[17]

f: *GW* pseudopotentials, basis set cutoff = 700 eV (else PBE pseudopotentials and 520 eV cutoff).

## 3.2 PES overview

In Fig. 2, the PES obtained using computational methods that include and exclude dispersion interactions are shown separately, with all results pertaining to a 6 × 6 × 4 *k*-point grid. A significant qualitative feature is the universally predicted stabilization of the hexagonal planar structure **C** of $Zn_{0.5}Mg_{0.5}O$ compared to that in ZnO. This feature has previously been identified as central to the enhanced ferroelectric behavior of the substituted material.[17]

For $Zn_{0.5}Mg_{0.5}O$, all methods predict that both the wurtzite form **A** and the planar hexagonal form **C** form energetically accessible polymorphs, with clear intermediate transition states **B** (Table 1, Fig. 2). The dispersion force is found to strongly stabilize **C** relative to **A**, with HF/D3, PBESol/D3, PBE0/D3, and HSE06/D3 all predicting **C** to be the most stable polymorph. This is inconsistent with experiment as only **A** has been observed, exhibiting ferroelectric switching through what is believed to be a **C** intermediate.[22,23] The r$^2$SCAN/rVV10 calculations instead predict **A** to be slightly more stable than **C** by 16 meV, whereas the MP2 and RPA calculations predict **C** to be significantly more stable by 59 – 74 meV (using the same k-point grid). These results highlight the considerable challenge faced by DFT approaches when considering the reactivity of $Zn_{0.5}Mg_{0.5}O$, that HF and especially HF/D3 can deviate significantly from higher level methods, and that the treatment of dispersion interactions is critically important to the portrayal of ferroelectric switching.

For the overall shape of the ZnO PES, the results are quite different. All DFT methods yield similar PES, whether or not dispersion interactions are included. The reduced influence of dispersion occurs owing to two factors: (i) the dispersion-induced stabilization of **C** is smaller, decreasing from 50-60 meV for $Zn_{0.5}Mg_{0.5}O$ to 30-50 meV in ZnO, and (ii) the large intrinsic energy difference between **C** and **A**, $E_C - E_A$. The MP2 and RPA results for ZnO are quantitatively similar and show good qualitative agreement with DFT calculations performed without inclusion of dispersion, with PBE0 and HSE06 giving the best quantitative agreement with MP2 and RPA. For both materials, the PBE0 and HSE06 results are nearly indistinguishable. A similar correspondence is observed for PBESol and r$^2$SCAN for ZnO, however, PBESol performs poorly for $Zn_{0.5}Mg_{0.5}O$.

This good qualitative agreement between methods is insufficient for ZnO, however, as a critical subtle issue concerns the nature of **C** as either a very shallow intermediate metastable phase or else as a transition state for concerted polarization switching. Without dispersion corrections, all DFT methods predict **C** to be a transition state, with the energy decreasing by 2 – 3 meV at $RC = \pm 1/8$, in agreement with the predictions of MP2 and RPA. Dispersion corrections stabilize **C**, significantly flattening the shape of the PES in in its vicinity, with the effect being large enough for PBE0/D3 and HSE06/D3 to change **C** into a metastable phase and introduce new transition states **B** at $RC = 1/16 – 1/8$ with barrier heights < 1 meV. For this key aspect of the switching mechanism, the treatment of dispersion is therefore again seen to be critical to understanding.

In experiments, **C** has been reported once as a metastable phase in ZnO nanocrystals that transforms to **A** above 200 °C.[24] This result suggests that **C** could be metastable in infinite crystals as modelled by the PBE0/D3 and HSE06/D3 calculations, but alternatively, the observed nanocrystals may be stabilized by finite-size or solvation effects. Discriminating between these possibilities requires careful assessment of the accuracy and limitations of the computational approaches applied to infinite crystals.

### 3.3 Lattice-vector comparisons

Insight into the qualities of the computational approaches can be obtained by examining the optimized lattice parameters for the **A**, **B**, and **C** structures. Table 2 shows the predicted values of $a$ whereas Table 3 shows those of $c$. In most cases, the spread in the calculated lattice parameters is of the same order as the variation in the available experimental data, indicating that all methods capture the essential structural features relevant to concerted ferroelectric switching in both materials. The average absolute differences between calculated and experimental lattice parameters (see Tables 2 and 3) range from 0.008 Å to 0.016 Å for PBE0, HSE06, and r$^2$SCAN, or else from 0.014 Å to 0.026 Å for PBE0/D3, HSE06/D3, and r$^2$SCAN/rVV10; the only clearly poor results were obtained by HF/D3, with deviations of up to 0.066 Å. The differences are most significant for $c$, but nevertheless are relatively small given the significant contraction that occurs in this direction as a function of $RC$, and are also small compared to the differences that, in the past, have challenged experimental data interpretations.[47] In summary, although the treatment of dispersion is shown to be critical to the understanding of energetics of polarization switching (Table 1), it plays only a minor role in predicting equilibrium lattice parameters and slightly worsens agreement with experiment.

**Table 2.** Calculated *a* lattice vector length for ZnMO$_2$ unit cells, in Å, for wurtzite structures (**A**), planar hexagonal structures (**C**), and interconnecting transition states (**B**), obtained with (w D) and without (w/o D) inclusion of dispersion.[a]

| M | Method | A[b] | | B | | C[c] | |
|---|---|---|---|---|---|---|---|
| | | w/o D | w D | w/o D | w D | w/o D | w D |
| Zn | HF | 3.289 | 3.209 | *3.439* | 3.334 | 3.455[d] | 3.386 |
| | PBESol | 3.240 | 3.221 | *3.415* | *3.404* | 3.426 | 3.412 |
| | r$^2$SCAN | 3.250 | 3.241 | *3.421* | *3.417* | 3.432 | 3.427 |
| | PBE0 | 3.258 | 3.236 | *3.429* | 3.422 | 3.440 | 3.423 |
| | HSE06 | 3.261 | 3.237 | *3.433* | 3.419 | 3.444[e] | 3.427 |
| Mg | HF | 3.269 | 3.194 | 3.437 | 3.293 | 3.474 | 3.370 |
| | PBESol | 3.268 | 3.251 | 3.389 | 3.359 | 3.477 | 3.454 |
| | r$^2$SCAN | 3.282 | 3.275 | 3.419 | 3.395 | 3.490 | 3.483 |
| | PBE0 | 3.267 | 3.245 | 3.403 | 3.370 | 3.477 | 3.449 |
| | HSE06 | 3.269 | 3.245 | 3.405 | 3.371 | 3.479 | 3.450 |

a: Either: D3(BJ) for HF, PBESol, PBE0, and HSE06; or rVV10 for r$^2$SCAN. When no intermediary structure is identified, the value at $RC = \pm 1/8$ is shown in italics.
b: Observed: for w-ZnO, 3.2493 Å[13] and 3.254 Å[110] for Zn$_{0.46}$Mg$_{0.54}$O, Å ~3.252[23] and ~3.284.[110]
c: Observed 3.45±0.02 Å (h-ZnO).[47]
d: Previously calculation 3.48 Å.[42]
e: Previously calculation 3.425 Å.[40]

**Table 3.** Calculated $c$ lattice parameters for ZnMO$_2$ unit cells, in Å, for wurtzite structures (**A**), planar hexagonal structures (**C**), and interconnecting transition states (**B**), obtained with (w D) and without (w/o D) inclusion of dispersion.[a]

| M | Method | A[b] | | B | | C[c] | |
|---|---|---|---|---|---|---|---|
| | | w/o D | w D | w/o D | w D | w/o D | w D |
| Zn | HF | 5.255 | 5.051 | *4.616* | 4.532 | 4.542[d] | 4.318 |
| | PBESol | 5.225 | 5.194 | *4.520* | *4.460* | 4.477 | 4.430 |
| | r$^2$SCAN | 5.235 | 5.219 | *4.546* | *4.508* | 4.498 | 4.471 |
| | PBE0 | 5.239 | 5.201 | *4.537* | 4.437 | 4.490 | 4.430 |
| | HSE06 | 5.240 | 5.198 | *4.529* | 4.447 | 4.486[e] | 4.420 |
| Mg | HF | 5.172 | 4.912 | 4.455 | 4.496 | 4.313 | 4.175 |
| | PBESol | 5.129 | 5.081 | 4.629 | 4.634 | 4.276 | 4.246 |
| | r$^2$SCAN | 5.159 | 5.147 | 4.586 | 4.649 | 4.301 | 4.291 |
| | PBE0 | 5.150 | 5.097 | 4.576 | 4.570 | 4.284 | 4.246 |
| | HSE06 | 5.152 | 5.094 | 4.574 | 4.562 | 4.283 | 4.241 |

a: Either: D3(BJ) for HF, PBESol, PBE0, and HSE06; or rVV10 for r$^2$SCAN. When no intermediate structure is identified, the value at $RC = \pm 1/8$ is shown in italics.

b: Observed: for w-ZnO, 5.2057 Å[13] and 5.202 Å[110] for Zn$_{0.46}$Mg$_{0.54}$O, ~5.188 Å[23] and ~5.11.Å

c: Observed 4.46±0.02 Å (h-ZnO).[47]

d: previously calculation 4.46 Å.[42]

e: previously calculation 4.512 Å.[40]

**3.4 Exploring other empirical dispersion corrections to DFT**

Given to the sensitivity of the energetics predictions for ZnO and $Zn_{0.5}Mg_{0.5}O$ to the treatment of dispersion, additional test calculations were performed using alternative dispersion corrections available in VASP. These calculations employed the structures **A**, **B**, and **C** previously optimized using DFT/D3(BJ) with either PBE0 or HSE06. Calculations using the dDsC correction,[86,87] as well as the TS[88] correction and its developments,[89-94] did not converge and were therefore excluded. In contrast, calculations completed successfully for D2,[95] D3 without Becke-Johnson damping,[96] and ULG correction.[97] These results are summarized in Table 4, where they are compared to D3(BJ), rVV10, MP2 and RPA results. Results are also included for the optB88 functional[111] that implicitly includes dispersion.

Overall, the best agreement between dispersion-corrected DFT approaches and the MP2 or RPA results is found for the optB88 and ULG approaches (see SI for average absolute differences). Next comes $r^2$SCAN/rVV10, followed by D2 and D3(BJ), with the poorest agreement observed for D3. In particular, the D2 and D3 methods show significant variations with the geometrical structure, suggesting limited reliability for the systems investigated.

Comparing all results obtained with and without dispersion, the average absolute deviations from MP2 and RPA are similar for PBE0, HSE06, PBE0/ULG, HSE06/ULG, and optB88, being approximately 8 – 10 meV (see SI). This implies that the ULG dispersion correction has minimal effect, and indeed its average contribution to $E_C - E_A$ and $E_B - E_A$ is just -11 meV (see SI). The effect is also small for rVV10 at -19 meV, but greater for D3(BJ) at -47 meV.

**Table 4.** Calculated structural energy differences (in meV per ZnMO$_2$ cell) resulting primarily from variations in the method used to treat dispersion interactions, obtained using different geometrical (Geom) and electronic (El) crystal structures.[a]

| M | Geom. Str. | El. Str. | dispersion | $E_C - E_A$ | $E_B - E_A$ | $E_B - E_C$ |
|---|---|---|---|---|---|---|
| Zn | PBE0/D3(BJ) | PBE0 | ULG | 248 | 242 | -6 |
|  |  |  | D2 | 354 | 338 | -16 |
|  |  |  | D3 | 246 | 244 | -3 |
|  |  |  | D3(BJ) | 215 | 215 | 0.1 |
|  |  | optB88 | intrinsic | 258 | 255 | -3 |
|  | HSE06/D3(BJ) | HSE06 | ULG | 246 | 245 | -1 |
|  |  |  | D2 | 298 | 295 | -3 |
|  |  |  | D3 | 74 | 76 | 2 |
|  |  |  | D3(BJ) | 206 | 206 | 0.7 |
|  |  | optB88 | intrinsic | 256 | 253 | -3 |
|  | r$^2$SCAN/rVV10 | r$^2$SCAN | rVV10 | 216 | 215 | -1 |
|  | PBE0 | HF | MP2 | 245 | 242 | -3 |
|  |  | PBE | RPA | 248 | 246 | -2 |
| Mg | PBE0/D3(BJ) | PBE0 | ULG | 40 | 93 | 52 |
|  |  |  | D2 | 105 | 116 | 11 |
|  |  |  | D3 | 6 | 83 | 78 |
|  |  |  | D3(BJ) | -5 | 68 | 73 |
|  |  |  | optB88 | 54 | 102 | 48 |
|  | HSE06/D3(BJ) | HSE06 | ULG | 41 | 93 | 53 |
|  |  |  | D2 | 77 | 109 | 32 |
|  |  |  | D3 | -51 | 19 | 69 |
|  |  |  | D3(BJ) | -14 | 63 | 78 |
|  |  |  | optB88 | 54 | 102 | 48 |
|  | r$^2$SCAN/rVV10 | r$^2$SCAN | rVV10 | 16 | 77 | 61 |
|  | PBE0 | HF | MP2 | 69 | 105 | 37 |
|  |  | PBE | RPA | 59 | 99 | 41 |

a: All calculations use a 6 × 6 × 4 $k$-point grid; the r$^2$SCAN/rVV10, MP2, RPA, and D3(BJ) results are taken from Table 1.

### 3.5 The energy of C at the lattice vectors of A

Naively, a significant effect for dispersion in the ferroelectric switching processes is expected owing to the chemical nature of the planar-hexagonal structure constrained to the lattice vectors of **A**, $C_v$, which adopts trigonal planar coordination characteristic of classic van-der-Walls bonded systems (Fig. 1). This effect is nevertheless largely offset by the significant lattice-volume change that occurs when $C_v$ collapses to the optimized trigonal-bipyramidal structure, **C**. To examine the effect of dispersion during this vertical lattice-vector transition, the energy differences $E_{C_v} - E_A$ are also listed in Table 1. As expected, all methods that do not include dispersion yield poor results compared to the MP2 and RPA ones. The best results are produced by PBE0/D3(BJ) and HSE06/D3(BJ) for both materials, with PBESol/D3(BJ) performing well for ZnO and HF/D3(BJ) performing well for $Zn_{0.5}Mg_{0.5}$. Hence, the importance of dispersion for ferroelectric switching depends on the specific pathway by which switching occurs, a feature that could become more pronounced if polarization reversal proceeds via a non-concerted mechanism.

## 4. DISCUSSION

Concerning computational methods, the central issue concerns the ability of different approaches to make quantitatively or qualitatively reliable predictions of relevant properties. Historically computational studies have contributed, e.g., by inspiring the reassignment of critical experimental data[47] for h-ZnO and by explaining of the effects of Mg substitution in promoting ferroelectricity.[17] Nevertheless, for quantitative predictions including subtle features such as the stability of h-ZnO, the question remains as to which are the most reliable computational methods.

From Table 1, the MP2 correction to the HF $E_C$ - $E_A$ energy difference is 62 meV for ZnO and 85 meV for $Zn_{0.5}Mg_{0.5}O$. This correction is modest for ZnO as it is just 20% of the total energy difference, but it becomes 55% for $Zn_{0.5}Mg_{0.5}O$ as the energy scale is much smaller. Nevertheless, the HF description is qualitatively reasonable, suggesting that the MP2 correction should be accurate. Moreover, the MP2 results are in good agreement with those from RPA, and hence two fundamentally different high-level methods portray the same qualitative scenarios: h-ZnO should be a transition state in the infinite-crystal limit, whereas h-$Zn_{0.5}Mg_{0.5}O$ should be a metastable polymorph of w-$Zn_{0.5}Mg_{0.5}O$.

Further quantitative calculation refinements confirm these conclusions. Increasing the RPA $k$-point grid from 6 × 6 × 4 to 12 x 12 x 8 delivers at most 3 meV change in relative energies (Table 1). Also reported are calculations using 6 × 6 × 4 $k$-points with a larger plane-wave cutoff of 700 eV and *GW* pseudopotentials. These changes made no difference for ZnO, but stabilized **C** relative to **A** by 17 meV for $Zn_{0.5}Mg_{0.5}O$. Given the small energy scale for ferroelectric switching in $Zn_{0.5}Mg_{0.5}O$, this results suggests that the improved treatment of Mg embodied in the *GW* pseudopotentials is important for achieving quantitative accuracy.

For ZnO, the PBESol, PBE0, HSE06, and r²SCAN methods all predict PES in close agreement with MP2 and RPA (Fig. 1), with $E_C$ - $E_A$ differences within ± 10 meV (Table 1). In contract, dispersion corrections based on D3(BJ) or rVV10 tend to over-stabilize **A**, with differences ranging from -45 meV to -30 meV (Table 1), although the ULG correction and optB88 perform well (Table 4). In summary, the conclusion is that all modern DFT methods overall provide a reliable description of ZnO energetics, and dispersion corrections are generally unnecessary, except when addressing subtle questions such as the nature of h-ZnO. Nevertheless, for $\mathbf{C_v}$ (structure **C** at the lattice parameters of **A**), dispersion corrections are required as all uncorrected methods lead to large differences in the range of -50 meV to -40 meV (Table 1).

For $Zn_{0.5}Mg_{0.5}O$, the situation is very different. Even though MP2 and RPA results are again similar and should provide reasonable approximations to the exact PES, the DFT results with and without dispersion spread over a wide range and systematically over-stabilize **C**. Inclusion of D3(BJ) or rVV10 makes the situation worse, to the point that only r²SCAN/rVV10 predicts **A** to be more stable than **C**.

The critical feature here is that only MP2 and RPA provide descriptions of the electronic kinetic energy, asymptotic potential, short-range correlation, and dispersion correlation that are self-consistent when comparing calculations of ZnO to those for $Zn_{0.5}Mg_{0.5}O$. This consistency arises from their seamless incorporation of both short- and long-range electron correlation. Hence it is most likely that the MP2 and RPA calculations provide the most reliable descriptions of h-ZnO, and that therefore the experimentally observed stabilization of h-ZnO nanocrystals may be due to solvent or finite-size effects.

## 5. CONCLUSIONS

The calculated PES for ferroelectric switching in ZnO and $Zn_{0.5}Mg_{0.5}O$ reflect competing effects that need to be balanced to obtain reliable predictions. For ZnO, this is achieved by essentially all computational methods considered (except HF/D3(BJ)), whereas for $Zn_{0.5}Mg_{0.5}O$ it is only achieved by MP2 and RPA. MP2 and RPA, together with most dispersion-corrected DFT calculations, indicate that the planar hexagonal phase **C** of ZnO (h-ZnO) should be unstable in bulk crystals. This suggests that the observed metastable nanoparticles were stabilized by surface interactions. For $Zn_{0.5}Mg_{0.5}O$, however, ambiguities in the performance of empirical dispersion corrections lead to scattered DFT predictions. Only MP2 and RPA, with possibly the exception of $r^2$SCAN/rVV10, predict that the wurtzite phase **A** is more stable than the conceivable planar hexagonal structure.

As only hard ions are present in these materials, the HF method provides a useful starting point for the description of ferroelectric switching, but it neglects electron correlation. This leads at short range to an underestimation of covalent effects (with associated overestimation of ionic effects), and at the long range to the absence of dispersion effects. MP2 and RPA correct all these shortcomings in controlled ways. As the resulting corrections are small, both of these approaches are expected to be quantitatively reliable. DFT functionals substantially improve upon the HF treatment of ionic bonding and short-range correlation, but for the hard-ion systems considered, empirical dispersion corrections fail to provide a cohesive qualitative description of dispersion. In addition, the methods used poorly represent the asymptotic potential and hence are unreliable for charge transfer, and present varying approximations to the electronic kinetic-energy operator and its effects on covalency.

Although the dispersion energy is established as being a significant contributor to concerted ferroelectric switching in these materials, its similar effect predicted for both ZnO and Zn$_{0.5}$Mg$_{0.5}$O implies that it cannot explain on its own the significant reduction in the switching barrier predicted, and observed, upon Mg substitution into ZnO. Clearly, the effects of this substitution on ionic and covalent bonding are perceived differently by HF and by the PBESol, PBE0, HSE06, and r$^2$SCAN functionals. Further in-depth analysis is therefore required to identify the origin of these changes and hence clarify the relative advantages and disadvantages of each computational method.

## ASSOCIATED CONTENT

**Supporting Information**. An Excel file is available that contains automatically generated summaries of each calculation performed, with the generated data collected to produce the results tables and figures, plus other material described in the main text.

## AUTHOR INFORMATION

**Corresponding Author**

* jeffrey.reimers@uts.edu.au; renwei@shu.edu.cn.

## ACKNOWLEDGMENT

We thank the National Natural Science Foundation of China (52130204, 12404276, 12347164), the China Postdoctoral Science Foundation (2024T170541, GZC20231535), the China Scholarship Council (No.202406890093), and the Australian Research Council (DE220101147, DP240103127 and Centre of Excellence in Quantum Biotechnology, CE230100021). We thank

TOC graphic

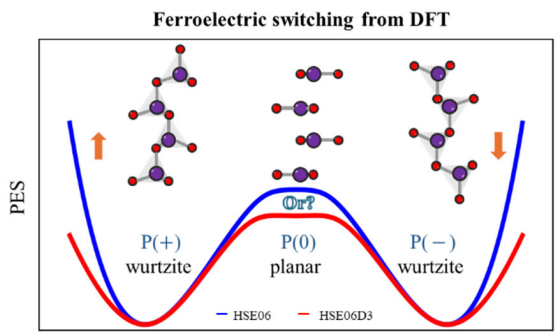